\documentclass[runningheads]{llncs}

\usepackage[T1]{fontenc}
\usepackage[breaklinks]{hyperref}
\usepackage{hyperref}
\hypersetup{breaklinks=true}
\usepackage{xurl}
\usepackage{xcolor}
\usepackage{lscape}
\usepackage[capposition=top]{floatrow}
\hypersetup{
    colorlinks,
    linkcolor={blue!80!black},
    citecolor={blue!80!black},
    urlcolor={blue!80!black}
}
\usepackage[sorting=none, style=numeric-comp]{biblatex}
\bibliography{references}
\usepackage{soul}
\usepackage[ruled,vlined]{algorithm2e}
\usepackage[affil-it]{authblk}
\usepackage{graphicx}
\begin{document}
\title{The Complex Network Patterns of Human Migration at Different Geographical Scales: Network Science meets Regression Analysis}
%
%\titlerunning{Abbreviated paper title}
% If the paper title is too long for the running head, you can set
% an abbreviated paper title here
%
\author{Dino Pitoski\inst{1}\orcidID{0000-0003-0431-5352}\thanks{Corresponding author (email: \href{mailto:dino.pitoski@uniri.hr}{dino.pitoski@uniri.hr})} Ana Meštrović\inst{1,2}\orcidID{0000-0001-9513-9467} \and
Hans Schmeets\inst{3,4}\orcidID{0000-0001-8511-723X}
}
\authorrunning{D. Pitoski et al.}
% First names are abbreviated in the running head.
% If there are more than two authors, 'et al.' is used.
%
\institute{Center for Artificial Intelligence and Cybersecurity, University of Rijeka, Croatia  \and 
Faculty of Informatics and Digital Technologies, University of Rijeka, Croatia \and Faculty of Arts and Social Sciences, Maastricht University, The Netherlands \and Department of Living Conditions and Social Cohesion, Central Bureau of Statistics of the Netherlands, Office Heerlen, The Netherlands
}
\titlerunning{Complex network patterns of human migration}

\maketitle              % typeset the header of the contribution
\begin{abstract}

Migration's influence in shaping population dynamics in times of impending climate and population crises exposes its crucial role in upholding societal cohesion. As migration impacts virtually all aspects of life, it continues to require attention across scientific disciplines. This study delves into two distinctive substrates of Migration Studies: the "why" substrate, which deals with identifying the factors driving migration relying primarily on regression modeling, encompassing economic, demographic, geographic, cultural, political, and other variables; and the "how" substrate, which focuses on identifying migration flows and patterns, drawing from Network Science tools and visualization techniques to depict complex migration networks. Despite the growing percentage of Network Science studies in migration, the explanations of the identified network traits remain very scarce, highlighting the detachment between the two research substrates. Our study includes real-world network analyses of human migration across different geographical levels: city, country, and global. We examine inter-district migration in Vienna at the city level, review internal migration networks in Austria and Croatia at the country level, and analyze migration exchange between Croatia and the world at the global level. By comparing network structures, we demonstrate how distinct network traits impact regression modeling. This work not only uncovers migration network patterns in previously unexplored areas but also presents a comprehensive overview of recent research, highlighting gaps in each field and their interconnectedness. Our contribution offers suggestions for integrating both fields to enhance methodological rigor and support future research.

\keywords{network science \and human migration \and spatial analysis \and statistical methods}

\end{abstract}
\section{Introduction}
\label{Sec1}

It may seem that the topic of human migration came to the forefront of political, mediatic, public, or scientific discourse only recently, with the severe displacement of Syrians in 2015 and the ongoing ``migration crisis'', exacerbated by the war in Ukraine and the new refugee fluxes. Yet, migration is an inherent aspect of human existence, that always has, and always will be of consequence for societal well-being. Whether it's internal migration within a country, or international migration between countries, the movement of people, coupled with natural population changes through births and deaths, shapes the population dynamics at various geographical scales. These changes can have a severe impact on the societal order, that should be maintained to ensure a harmonious and stable life for individuals, and the overall stability of societies. 

There are strong predictions that the ``migration crisis'' is to be exacerbated by the ``climate crisis'' that works in parallel, and intensifies \cite{clement2021groundswell}, so the social cohesion issues might be expected to become even more mainstream. Moreover, the forthcoming ``population crisis'', which is a result of two thirds of all people living in countries where the average birth rates are lower than the replacement rate, brightens the spotlight for migration as an important part of a solution \cite{UNPF2023}. 

As being such a consequential matter, migration became the topic of inquiry across an increasing number of scientific disciplines, whose output (studies published as books, research papers, policy reports, and increasingly online platforms) are so vast in number that it has called for the development of a migration research database; the so-called ``Migration Research Hub'', available at \url{https://migrationresearch.com/}. Established under the auspices of the European Union, the database has the purpose of aggregating, maintaining and sorting the global migration research output in one place, for the policymakers and researchers to be able to keep track on the evidence on migration as basis for their evidence-informed policies and research. While initially perceived to fall solely under the realm of Social Sciences, ``Migration Studies'' have evolved into a new interdisciplinary scientific field that encompasses a wide range of disciplines and methodologies \cite{Pisarevskayaetal2020}.

In this study, we focus on two identified distinct substrates of Migration Studies. The first, the so-called \textit{``why''} substrate, essentially deals with the identification of factors (often termed as ``drivers'') of human migration, taking migration flows or migrant stocks as dependent variables, and quantifying the influence of independent variables on the generation of these flows. The independent variables for the models are pulled out from various baskets: economic (such as income and wages, GDP per capita, (un)employment rates, costs of living or moving), demographic (such as age, gender, population size, density or growth, household or marital status, education levels), geographic (such as physical distance, proximity or contiguity), cultural (such as language proximity, colonial relationships, religion), political (such as political freedoms or migration policies' restrictiveness), and various other (such as the increasingly important environmental factors of floods and droughts, temperature and precipitation, natural disasters, various sorts of inequalities, levels of poverty, violence, etc.). All in all, a vast number of wider contextual, as well as individual factors, get evaluated for their (strength of) effect in driving or deterring migration across the studied geographies. The primary evidencing method of these influences shows to be regression modelling, often promoted as the most reliable, as being based on quantitative data and rigorous inferential methodology. 

The second, \textit{``how''} substrate, focuses at telling how migration takes place, that is, at migration \textit{flows}, migration \textit{patterns}, or, ultimately, \textit{networks} of human migration flows. This substrate ties to the emerging field of Network Science, which increasingly penetrates into Migration Studies (as it does into virtually all domains). Network science (NS) deploys social network analysis tools: indicators, algorithms, models, and visualizations, to depict human migration flow patterns abstracted as complex networks from, again, quantitative data. Through these visualizations and the accompanying analytics, NS tries to deliver the comprehensive view on the migration network behaviour. This allows the social demographers, economists, and other experts (from the ``why'' substrate) to relatively easily identify the potential reasons underlying the discovered network patterns \cite{Schon2021, Pitoskietal2021c}, hence, reduces complexity when it comes to selecting potential variables to deploy in, e.g., regression-based explorations. However, although the number of Network Science studies on human migration appears to be quickly rising in terms of percentage growth, the number of studies (hence, the number of investigated territories) is still very small, while the network analyses performed do not contain an epilogue, in the sense of providing substantial explanations on the factors that underlie the traced network appearances. The network analyses essentially stop at providing the answers to the question \textit{how} network patterns look like, and the two substrates remain detached from each other. 

In the sequel, we present the real-world network analyses of human migration, spanning geographical scales from city to country to global level. For the city level, we provide a new empirical network analysis of inter-district migration in Vienna, Austria. At the country level, we provide a brief review of two recent NS-based studies on internal migration: one focusing on Austria \cite{Pitoskietal2021a} and the other on Croatia \cite{Pitoskietal2021b}. For the global level, we provide another new empirical network analysis of the migration exchange of Croatia with the rest of the world. We compare migration network features identified at the three different levels and show how some pertinent network traits affect the deployment of regression models. Apart from offering the insight into migration network patterns in yet uninvestigated territories, we provide a thorough overview of the up-to-date migration research output in the two previously addressed strands of research. We expose the gaps present in each individual strand, as well as the gaps that one reveals about the other. Our ultimate contribution is the set of substantiated propositions on how to deploy both fields simultaneously as to achieve methodological soundness, ``making the most'' from both, and aiding to future research. 

In the next section, we present the related work and the status quo in both research domains, along with their individual gaps. In Section \ref{Sec3} we examine migration-network patterns in the three aforementioned geographies. We thoroughly describe these networks' features and provide comparisons between them with regards to the different geographical spans observed, with emphasis to the geographic validity of findings about their behaviour, or network consistency across these different levels. In Section \ref{Sec4} we discuss the consequences of the findings, outlining the potential problems when two strands of research, their methods in particular, are not properly combined. We close the same section by summarizing the main points, and offering our ways for improvement.  

\section{Related Work}
\label{Sec2}

\subsubsection{Migration Factors Studies.}

The factors driving human migration are examined over a massive amount of studies attaching to various migration-theoretical frameworks, such as the push-and-pull theory \cite{Lee1966}, theory of migration systems \cite{Mabogunje1970}, migration or mobility transition frameworks \cite{Zelinsky1971, Skeldon1990}, neo-classical migration theory \cite{HarrisandTodaro1970}, dual labour-market theory \cite{Piore1979}, new economics of labour migration \cite{Stark1978, Stark1991}, theory of cumulative causation \cite{Massey1990}, or the recent aspirations-capabilities framework \cite{deHaas2021}. At the ``peak of congestion'' in Migration Studies, where both the developed theories and the range of factors traced as probable determinants of migration have been accummulating to the point when systematization is desperately needed (see the points raised in \ref{Sec1}), two valuable systematic reviews recently emerged. These provide a summary of the determinants of \cite{Aslanyetal2021} \textit{migration aspirations} utilized as the dependent variable in survey-based literature  (49 studies using data from 1990 onward), and \cite{Pitoskietal2021d}  of the \textit{realized migration} as the dependent variable\footnote{The model types, the operationalization of the migration as the dependent variable, as well as migration factors as predictors, varies across literature. The multitude of definitions and means of measurement have been listed in the Supplementary Material of \cite{Pitoskietal2021d}.} used in mainly regression-model- based literature (163 studies using data from 1990 onward). The studies were not limited to reviewing factors of international migration with countries as a whole as origins and destinations, but also include internal migration factors and their influence when it comes to migration between human settlements within countries (cities, towns and villages). Comprehensive visualizations that provide the insights on the established relevance of particular migration aspirations are available as figures in the aforementioned \cite{Aslanyetal2021}, and the interactive Migration Drivers Map on the factors of realized migration working at different geographies, developed as part of \cite{Pitoskietal2021d}, is available at \url{https://tabsoft.co/3VzqXux}.

The summary of aggregated findings of these works are as follows. The main determinants of migration aspirations that work as \textit{push factors from the origins}\footnote{In this representation, we reduce to the push-and-pull theory \cite{Lee1966} terminology for the simplicity of language.}  are found to be: violence and insecurity, previous migratory experience or belongingness to migrant communities (the so called ``migration networks'', not to be mixed with the graph-theoretical concept of migration networks discussed and analysed in this article), young age, male gender, urban residency, higher educational attainment, and larger household size. The determinants of migration aspirations that have a negative influence - i.e. that \textit{keep migrants at origins} are found to be: the sense of well being, employment, social attachment and participation, positive societal change over time, socio-economic status, being married or owning a home. 

As regards the determinants of actually realized migration, the main \textit{push factors from the origins} are found to be: educational attainment, unemployment, population size, previous migratory experience, male gender, and young age. The main \textit{pull factors toward destinations} have found to be: belongingness to migrant communities or having families abroad, income levels and economic development (most often represented by GDP), population size, and social protection expenditures. In addition, the ubiquitously proven \textit{deterring factor of the origin-destination migration exchange}, is geographical distance, while country/locational contiguity and language proximity are established as the \textit{spurring factors for the origin-destination migration exchange}. The latter deterring and spurring (i.e., \textit{intervening}) factors for the origin-destination migration exchange can be regarded as factors working at the \textit{link} level in network-scientific terms.

Migration factors may be fitted under few general categories if migrants would be questioned about their reasons for migrating \textit{at destinations}. The basic categories are arguably i) migration for work, ii) migration for education, iii) migration for family reunion, and iv) migration for seeking shelter from any kind of endangerment at the origin (e.g. asylum seekers and refugees). In addition, the reasons for migration can include those highly individualistic, such as the individual's quest for new experience, or the attraction to the amenities of a specific location. Some of the data on the reasons for individuals' reasons for actually migrating (the aforementioned basic categories) are gathered by some governmental institutions, and solely when it regards international migration (to the Netherlands, for example; see \cite{CBS2018}). Yet, such data, and thereby a clearer view on the balance between these factors for more geographies, is not available. For the Netherlands, it has been shown that the reasons are predominantly family reunion (around 50\% of international migrants), followed by work or education (each category pertaining to about 20\% of international migrants), and the rest (10\%) pertaining to asylum-seekers, for the period of 1999-2010 \cite{Schmeets2019}. Such efforts to collect the individual reasons or migration at destinations would complete the view on the relevance of factors that the scientific research has brought forward on migration aspirations, and the determinants of the realized migration.

Ultimately, what Migration Studies have managed to establish up to date is some approximation on the migration factor effects working on two scales, the international (inter-country) migration level and the internal (intra-country) level, basing these approximations predominantly on the realized migration, and predominantly on regression modelling. It is important to address here the fact that international migration has received much larger attention in migration factors research, although its size in terms of the number of migrants in global population is relatively small consistently over decades, at around 3.5\% (or 272 million persons, estimated in 2019). At the same time, internal migration is a about three times larger phenomenon (763 million persons, estimated in 2013) \cite{McKenzie2022}. Moreover, the data on the exact settlement-settlement relocations (i.e. specific city/town/village to specific city/town/village) are available only at the internal-migration level, while for the international migration the scientists are still able to work only with estimated country-level migration stocks \cite{WBmigdata} prompting the search for more accurate micro-level data sources for international migration \cite{Tjaden2021}. The lack-of-data issue impacts the reliability of regression models when international migration between countries is taken as a dependent variable in regression models with the aforementioned factors deployed as predictors, control or instrumental variables. This is, however, just one of the problems when regression analysis is concerned. The other issues relate to the particular traits of migration networks, that we identify through several real-case network analyses in the paper. These subsequently expose the potential methodological flaws in regression applications. At this point we continue with the review of network science in the migration domain, followed by the analyses, while we return to the methodological issues in large detail in the Discussion (\ref{Sec4}).

\subsubsection{Migration Networks Studies.}

A systematic literature review of studies that deployed the NS approach on human migration has recently been performed by \cite{Pitoskietal2021c}. The study identifies 22 of such works published by the beginning of 2019. The network abstraction utilized across these works is such where nodes are locations (i.e. countries, counties, cities, municipalities), and link weights (or simply existence) are determined by migration flows between locations, or, more commonly, stocks of people of citizenship of particular country in other countries. Most frequently used indicators and algorithms used for the analysis of migration networks are shown to be: (weighted) degree and other variations of node (location) centrality, transitivity including node and network clustering coefficients, network assortativity, network modularity and community detection algorithms based on modularity optimization. The same review also identifies network-geographical levels that have been researched, where the  predominant geography is the World (global international migration), European countries (intra-EU international migration) followed by the United States, China, UK and Mexico (countries' internal migration). From the beginning of 2019, at which time the coverage of the aforementioned review ends, a relatively large number of new network analyses has emerged, which is a demonstration of an increased involvement of Network Science in Migration Studies. 

Continuing with the same approach of \cite{Pitoskietal2021c} in terms of browsing of bibliographic databases to include new works in this summary (browsing ending with 30th May, 2023), we have identified the following: \cite{Bonaccorsietal2019},\cite{Aleskerovetal2020}, \cite{PoratandBenguigui2021}, and \cite{Akbari2021} analysed the global international migration network, \cite{Windzioetal2019} analysed international migration network of the countries of the European Union, \cite{GursoyandBadur2021} analysed the inter-state migration network of the United States, \cite{GursoyandBadur2022} analysed the internal migration network of Turkey,
\cite{CarvahloandChalresEdwards2020} analysed the internal migration network of Brazil and \cite{Chenetal2021} analysed the internal migration network of England and Wales. In addition, internal migration networks have been analysed in the two articles which we present in greater detail in Section \ref{Sec3}; that of Austria \cite{Pitoskietal2021a} and that of Croatia \cite{Pitoskietal2021b}. Our extended review has also identified new network analyses of (international) migration for specific population segments; sex workers \cite{Rochaetal2022}, refugees \cite{Mourao2020}, and innovators \cite{Wangetal2020}. A very brief overview of the findings from these works can be summarized as follows: there are consistent small-world features, high network clustering, and robust community structures traced in all of the observed networks, international or internal. In networks of international migration, developed countries act as migration attractors in the hierarchy of attractiveness between countries, with distinct rich-club patterns\footnote{Elaborating in more detail on the findings for each of the work would be infeasible in terms of increasing the length of the paper, thus we invite the reader to examine each of the specified work for the exact findings.}.

More consequentially, what can be found by examining all the new works is in line with the conclusions of the former \cite{Pitoskietal2021c} when addressing the greenness of the field; the same set of indicators and algorithms keeps being utilized for the analysis of migration networks. Very little emphasis still is put on how to viably abstract these networks from data for the reliable application of these indicators and algorithms. This initial step is, however, crucial for warranting any conclusions stemming from the employed analytics \cite{Pitoskietal2021c}. While intense weights on self loops has been confirmed for internal migration networks \cite{Pitoskietal2021a, Pitoskietal2021b}, this trait, along with another prominent trait, of (weighted) reciprocity on links, tends to be regularly disregarded when abstracting migration networks from available data. It has been demonstrated in these former works that migration on self loops and high reciprocity disturb the application of conventional network algorithms and indicators, and that in order for these indicators to be straightforwardly applied, the network abstractions need to be established in such a way that they lose the least of the information that they originally bear. For example, the removal of looping edges, the aggregation of reciprocated links into undirected weighted links, and similar operations should be avoided, and such procedures scrutinized. 

The optimal way for the Network Science in Migration Studies, as well as in other domains, to move forward is to start observing the \textit{space-time}, or \textit{temporal} - or \textit{dynamic} - network abstractions, to obtain more meaning from the indicators and algorithms applied. At the same time, the indicators and algorithms need to be upgraded to serve such abstractions, in the similar way as it recently has been offered by \cite{Pitoskietal2023a}. Moreover, an even more fine-grained network abstraction in which nodes would be space-time positions of particular individuals (see \cite{Pitoskietal2023b}). However, considering the type of data currently available for scientists to abstract migration networks from, \textit{static} weighted directed networks, with nodes as human settlements and links as within-period aggregated migration flows between them, currently is the only kind of abstraction that can feasibly be utilized, due to availability, but also due to accuracy and timeliness, of data being gathered and made available by collecting institutions \cite{Pitoskietal2021a}. This is how we also proceed in the analysis part of this article, commencing in next section.

\section{Network patterns of migration} 
\label{Sec3}

\subsection{Network data, abstractions, and tools for the analyses}
\label{gendef}

Section \ref{Sec3} as a whole provides the analyses of human migration network patterns at different geographical levels. We start off by analysing migration network at the narrow geographical span: a city and the migration between its districts. We then proceed to analyse migration networks on an intermediate span: migration network within countries, between their cities/municipalities. We end by analysing the networks on the widest geographical span: the international migration of people between cities/municipalities of a country and the rest of the World. These analyses are done respectively for Vienna, Austria and Croatia, based on the data acquired from each country's national statistical office. References to the exact sources are provided in each respective subsection. In this subsection we provide the generic formulation for all analysed networks in mathematical terms. 

The abstraction of the subsequently analysed migration networks from the available data is a weighted di-graph $\mathcal{G}=(\mathcal{N},\mathcal{L}, \mathcal{W})$, whose:

\begin{enumerate}
\itemsep=0pt
\item nodes $\mathcal{N}=\left\{n_1, n_2, ..., n_{N}\right\}$ are administrative subdivisional units: city districts, country cities or municipalities, and countries as whole, respectively of the observed geographical scale,
\item link weights $\mathcal{W}=\left\{ w_{ij} \right\}_{N \times N}$, $i,j=1,...,N$ (where $i$ can be equal to $j$) are the counts of official changes of address of residence from a city district, municipality or country (node) $i$ to a city district, municipality or country (node) $j$ in any given year, and
\item links $\mathcal{L}=\left\{ l_{ij} \right\}_{N \times N}$ is a binary projection of $\mathcal{W}$, such that $l_{ij}=1$ if $w_{ij} > 0$, and $l_{ij}=0$ if $w_{ij} = 0$.
\end{enumerate}

From this root abstraction, in which the self-loops are taken into account ($w_{ii} \geq 0$), we may reduce our observations to subgraphs $\mathcal{G}^{\prime}=(\mathcal{N},\mathcal{L}^{\prime}, \mathcal{W}^{\prime})$, where $\mathcal{W}^{\prime}=\mathcal{W} \setminus \left\{w_{ii}\right\}$ and $\mathcal{L}^{\prime}$ is the according binary projection of $\mathcal{W}^{\prime}$. 

In each case we observe \textit{static} weighted directed networks, in which the weights on links are represented by the aggregate yearly migration flows between node dyads respecting the direction of flows. As regards the network-analytical tools applied on these abstractions, these will be specified throughout each following case network analysis, but, overall, these comprise nodal indicators such as centrality measures (weighted degree, PageRank, HITS algorithm cenrality, etc.), and network-structural indicators, such as reciprocity, modularity, density, etc., including community detection algorithms. For the issues on data for the network abstractions and the choices on the metrics, we point again to the review of \cite{Pitoskietal2021c}.

\newpage
\subsection{Intra-city migration network: Vienna}
\label{Sec3.2}

This subsection is dedicated to the network analysis of within-Vienna migration as a city-level case analysis (most narrow geographical span). The data from which we abstracted the network come from the Austrian Bureau of Statistics, and are publicly available at at: \url{https://data.statistik.gv.at/web/catalog.jsp}, ``Population'' section. The data used for this specific analysis are those containing migration flows between/within city districts of Vienna and between/within municipalities and cities in Austria (``Wanderungen innerhalb Österreichs''), while we also used the polygon data related to the administrative division into Vienna's districts and Austrian municipalities (``Gliederung Österreichs in Gemeinden'') for the visualizations (deriving centroids of each polygon as geolocations). Specific years of observation are 2002, 2010, 2018 and 2019, and the multi-period observations were made in order to support the historic validity of the findings.

The main focus here is intra-Vienna migration, but we start off by providing some general insights on its size relative to internal migration in Austria as a whole. This may serve as an introduction to the analysis presented in the following subsection, in which we expand the coverage to internal migration between all Austrian cities and municipalities, while the reader can also obtain the general impression on the size of the migration phenomenon in the country as a whole.

\begin{figure}[ht]
    \centering
    \includegraphics[width=0.9\textwidth]{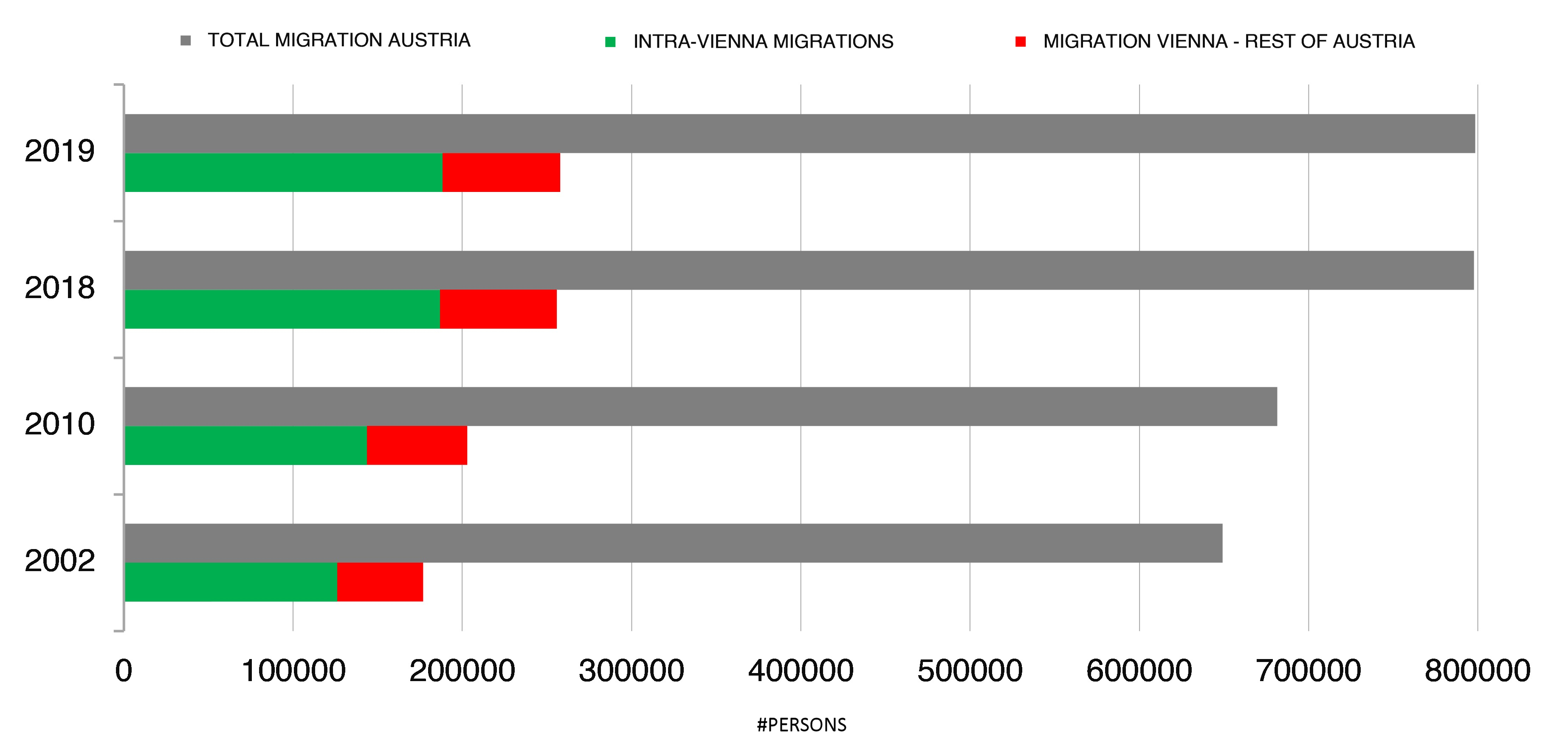}
    \caption{\textbf{Vienna-related migration in total Austrian internal migration, 2002-2019}}
    \label{fig:1}
\end{figure}

The data underlying the figure reveal that the size of ``Vienna-related'' migration is about 30\% of the size of the total internal migration in Austria; from that overall share, about 22\% pertains to exclusively intra-Vienna migrations, while about 8\% pertains to migration exchange of Vienna with the rest of Austria. Roughly about 800.000 internal migrations in the country as a whole have been noted in the last period observed (year 2019), while the country is populated by approximately 8 million people; hence, internal migration as the share of the country's population in general is about 10\%.\footnote{In this study we do not deal with the important connection of population size and migration per each city/country examination, but point to the thorough demonstrations on these connections provided in \cite{Pitoskietal2021a}}. Both the intra-Vienna migration and Vienna-rest-of-Austria migration grew relatively significantly throughout the observed period: 6.7\% and 3.6\% on average, respectively. This goes along with the generally increasing phenomenon of internal migration in the country, but Vienna-related migration increase seems to have been more rapid. 

\begin{figure}[ht!]
    \centering
    \includegraphics[width=\textwidth]{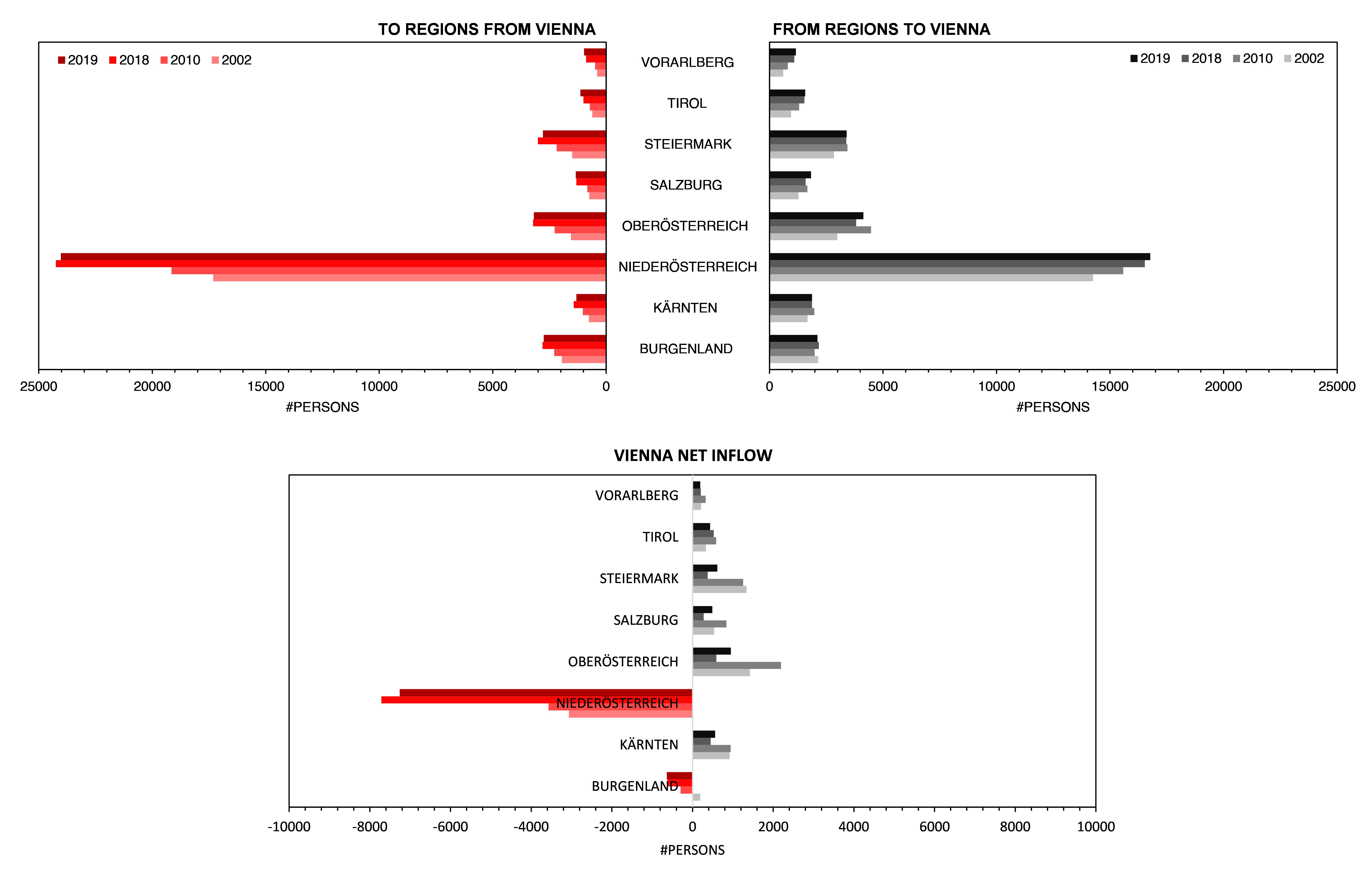}
    \caption{\textbf{Migration exchange of Vienna with Austrian regions, 2002-2019}}
    \label{fig:2}
\end{figure}

We further show, in Figure \ref{fig:2}, how Vienna connects with the rest of the country, addressing migration exchange with Austrian regions in particular. 

The exchange is exceptionally strong with the region of Niederösterreich (Lower Austria), which consistently receives more migrants from Vienna than it sends to Vienna. The most probable explanation of this behaviour is the geographical proximity of the region and its relatively large population size; the second largest in size with about 1.7 million people after Vienna itself (about 1.9 million people), based on the most recent population statistics \cite{migrationgvat_geography_population}. However, the exceptionally high volume of the non-reciprocated migration should attract researchers to thoroughly explore the drivers that underlie this and such exceptions for specific country examples. One of the rare studies focusing on Austria's internal migration drivers, \cite{Jestletal2022} highlights the role of relative deprivation, operationalized as income inequality. The study singles out Lower Austria, ranking third in income inequality within the country during 2011-2012, as a significant region for emigration. This to some extent conflicts our findings on the greater attraction of people towards Lower Austria consistently over the years, including the period covered in the aforementioned paper. This probably connects to the way migration has been operationalized as a dependent variable in the used model, the issue we return to in the Discussion (\ref{Sec4}). Additional insight that can be drawn from Figure \ref{fig:2} is that network reciprocity is a strong underlying feature of the network of migration, as there is a relatively steady flow-counterflow balance between Vienna and Austrian regions present over time.

\begin{figure}
    \centering
    \includegraphics[width=\textwidth]{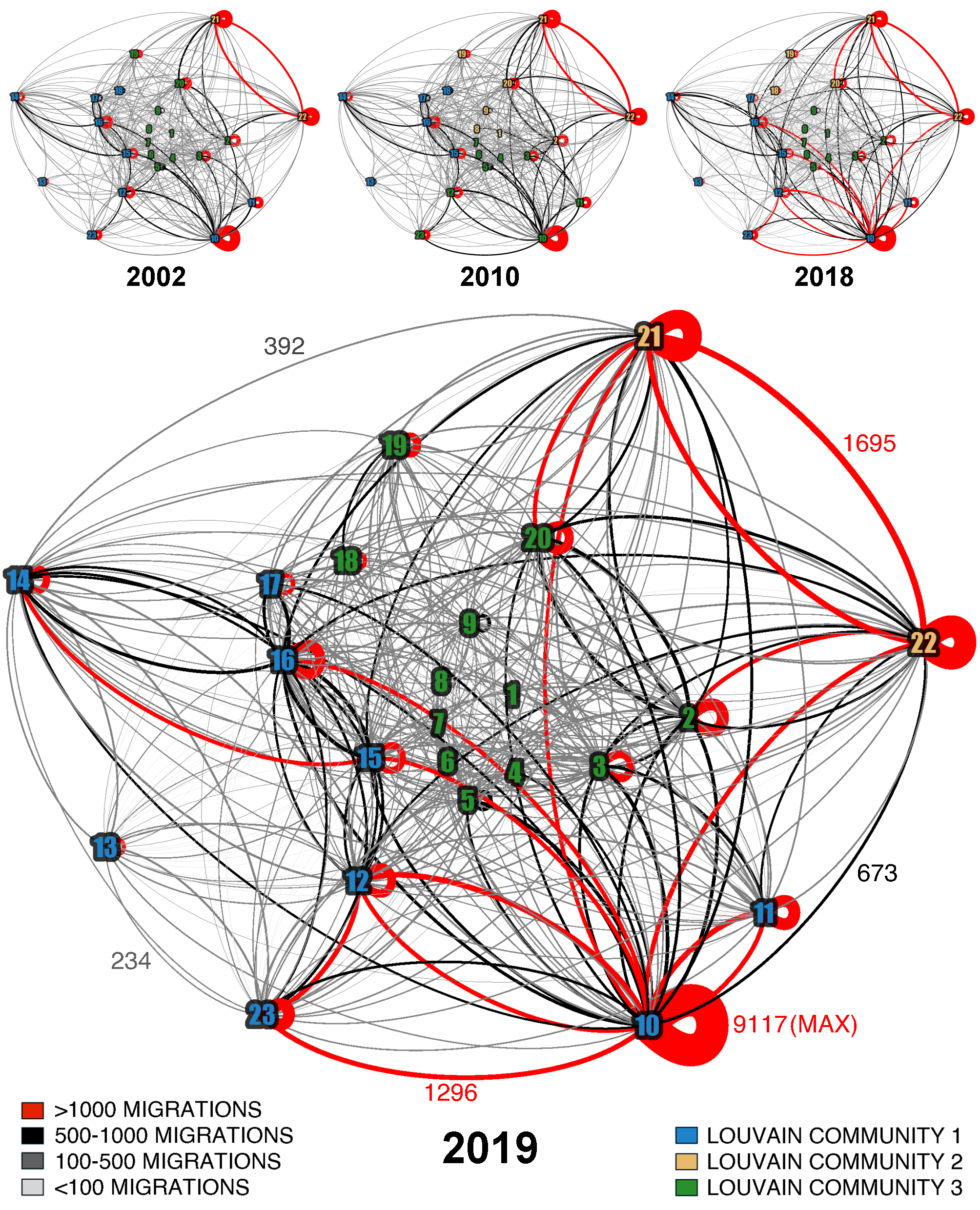}
\caption{\textbf{Evolution of the internal migration network of Vienna, 2002-2019}} 

\floatfoot{\small{Direction of~migration is represented by clockwise curvature of links. Link thickness proportional to the size of migration (see labeled edges for a general orientation). Communities detected by Louvain algorithm modularity optimization algorithm \cite{Blondel2008-ug}. District names: 1 - Innere Stadt, 2 - Leopoldstadt, 3 - Landstrasse, 4 - Wieden, 5 - Margareten, 6 - Mariahilf, 7 - Neubau, 8 - Josefstadt, 9 - Alsergrund, 10 - Favoriten, 11 - Simmering, 12 - Meidling, 13 - Hietzing, 14 - Penzing, 15 - Rudolfsheim-Fuenfhaus, 16 - Ottakring, 17 - Hernals, 18 - Waehring, 19 - Doebling, 20 - Brigittenau, 21 - Floridsdorf, 22 - Donaustadt, 23 - Liesing.}}
    \label{fig:3}
\end{figure}

A comprehensive (i.e., a network) view on the inter-district intra-Vienna migration flow patterns and their evolution over time, is provided in Figure \ref{fig:3}. As the graphs of the complete network ($\mathcal{G}$) clearly show, there is intense migration on self-loops (intra-district migration), while there is intense reciprocity on district dyads (inter-district migration). Total weights on self-loops comprise about 28\% of all intra-Vienna migration, per each of the observed periods except year 2002, when this percentage was much lower (about 16\%). This growth of ``looping'' migration appears to follow, with some lag, the growth of population in the city in general; namely, Vienna's growth in population compared to each next investigated period (population at years' end 2002 vs. 2010, 2010 vs. 2018, and 2018 vs. 2019), according to \cite{Magistrat-Stadt-Wien-2020}, was about 11\%, 9\% and 2.5\% respectively. As for the migration flows that are not distributed on self-loops, these are distributed on an established set of district dyads. The specific structure of the intra-Vienna migration network is ``fixed``; with time, the graphs don't seem to progress towards completeness, but the weight of migration tends to be distributed, and growing, on a unique set of links. 

Weighted reciprocity, measured as proposed by \cite{Squartini2013-xw} on $\mathcal{G}^{\prime}$ for the year 2018, is calculated at 0.758, which roughly means that for every 100 people moving from district A to district B of Vienna in 2018, there will be, on average, about 76 people moving from district B to district A in that same year. Year 2018 has been selected to match the analyses of internal migration in the country(s) as a whole, which follow in next subsection. Limiting to this one year only, we believe, does not fringe the historic validity of findings, especially given the fact that the values for the Pearson correlation coefficient between the weights on respective inter- and intra- district links in any two years compared (thus, including 2018) never has shown to fall below 0.98 in our calculations.

As final part of the intra-Vienna migration network analysis, embedded in Figure \ref{fig:3}, we ran the Louvain community detection algorithm \cite{Blondel2008-ug} on the undirected abstraction of $\mathcal{G}^{\prime}$, which exposed three distinct district communities: the districts in the South-West, those in the North-East, and those of the city centre. These specific communities continue to be more strongly integrated relative to the city network as a whole, consistently through the overall analysed period. A potential explanation of why we find these specific community formations is that the migrants' preferences for relocation are determined by the individuals' familiarity or some kind of a personal attachment to a specific region or part of the city.

\subsection{Intra-country migration networks: Austria and Croatia}

\subsubsection{Internal migration in Austria.} 

This, as the subsection that follows, reviews the previous NS studies that have been performed for migration in a country as a whole. Here we summarize the findings on the internal migration network with links (weights) being migration flows between all Austrian municipalities \cite{Pitoskietal2021a}, as an expanded view from the city to a country scale. The summary highlihgts only the concepts most related to the previous analysis section, for comparability. The reviewed study has thoroughly covered the network's evolution (each year from 2002 to 2018), explaining the inherent relationship of population and migration, and elaborating on the general patterns of migration, all while regarding the feasible network theoretical tools given the specific network structure. We invite the reader to assess that work for details and discussions on the calculated network metrics, as well as for the comprehensive network visualizations. An interactive visualization of the Austrian internal migration network in 2018 is available at: \url{http://bit.ly/3VnkYcQ}.

At the wider geographical scale (internal Austrian migration), extending from the narrower, city scale (intra-Vienna migration), the reviewed work shows some of the most prominent network features hold: the very high share of weights distributed on self-loops, as well as very high weighted reciprocity. Migration on self-loops took about 50\% to 55\% of all relocations in any year of observation; most of the migration at the country level is actually the relocation from one to the same city or municipality. Weighted reciprocity of inter-settlement relocations, was measured consistently at about 0.60 (60\%), which roughly means that for every 100 people moving from municipality A to municipality B in Austria, there will be, on average, about 60 people moving from municipality B to municipality A in the respective year of observation. We remind that the measured values respectively for the ``looping'' and reciprocal migration were measured at 28\% and 76\% or the intra-city (Vienna) network. The tendency of people to relocate within the one and the same district of a city appears to be lower at the smaller geographical scale, which seems natural as there might be less options to find a new living space in such close proximity to the existing. Yet, the high value of reciprocity and the localized distribution of communities, as outlined in the previous section, still reflects the people's tendency to relocate in and around ``familiar'' locations. In a sense, reciprocity too can be considered as looping migration within a confined geographical space. The addressed prominent network characteristics are shown for the case of Austria, and these appear to be similar at another country example, as in the review highlights of the study we provide next.

\subsubsection{Internal migration in Croatia.}
\label{Croint}

A network study comparable to the previously reviewed (Austria) that used the same set of network indicators, algorithms and visualizations drawn on the same kind of network abstraction (see Section \ref{gendef}), has been performed to analyse internal migration in Croatia \cite{Pitoskietal2021b}. By reviewing another country case we wanted to highlight the consistencies of the demonstrated internal migration network behaviour. The study on Croatia examines migration flows in the country in a single year (2018), based on the available data obtained from the Croatian Bureau of Statistics \cite{authorsown2}. The analysis also concentrates on the feasibility of network theoretical tools throughout their application, given the specific network structure examined. The reader is invited to assess the aforementioned study on Croatian inter-settlement network for the calculated network metrics and visualizations. An interactive visualization of the Croatian internal migration network in 2018 is available at: \url{http://bit.ly/3OK1tsg}.

What is specific for Croatia is that the total weights distributed on self-loops are found to be much lower than that in Austria; about 20\% of all migrations in that one year observed. Also, weighted reciprocity tends to be a bit lower, 49\%. In the same manner of explanation as before, we may say that roughly, for every 100 people moving from municipality A to municipality B in Croatia, there will be, on average, about half as many people moving from municipality B to municipality A, in the respective year of observation. When $\mathcal{G}^{\prime}$s of both countries for the respective period of 2018 were compared, a much stronger exchange was found between virtually all relatively more populated cities/municpalities and the capital city Zagreb. Less return flows to cities/municipalities were seen in specific regions, most prominently the region of Slavonia in the country's East, all of whose large cities send many more migrants to the country's capital (Zagreb) then they receive back from the same capital. These specifics may be explained by more business opportunities existing in the capital, which is inhabited by one third of the entire country's population (according to the recent census pursued in 2021 and published by the Croatian Bureau of Statistics \cite{dzs2023}). In addition, by the same census and from the new data which we present in detail in Subsection \ref{cromig}, it is clearly visible that Croatia has extremely high emigration rates (and from the country's east in particular). This perhaps also explains why the internal migration as a phenomenon in Croatia in general, as compared to that in Austria, was found to be much lower (around 2\% for 2018, as measured by the sum of total migration weights as the share of total population). 

Notwithstanding some of the precise country specifics, from both reviews it can be derived that the general migration network patterns at the country level are consistent with those at city level, with migration on self loops, and especially weighted reciprocity being their prominent characteristics. We believe it is very likely that one would find similar patterns if one would run the analysis on an intra-Zagreb (inter-district) migration network. However, we were not able to attain the data required to perform that analysis, as Croatian Bureau of Statistics, who were approached, do not maintain this data at the district level. Although we are aware that missing out city level comparisons may be perceived as a limitation, we believe this level of coverage should be satisfactory, considering that we trace similar patterns when widening the analysis to even larger spatial levels, as demonstrated in next section.

\newpage
\subsection{International migration network: Croatia and the World}
\label{cromig}

In this, final part of our analysis, we broaden our focus to the worldwide scale with the attempt to explain, and verify, whether migration network- structural features hold, or how much they differ, from the previously depicted on the smallest and the intermediate scale. We examine a new case network, comprising the migration flows between World countries and Croatian settlements (the latter represented by 554 Croatian cities and/or municipalities). Migration data, obtained upon request from the Croatian Bureau of Statistics, comprise the total number of immigrants (irrespective of subcategories), per each year in the period from 2016 to 2021, per Country citizenship of immigrants, and vice versa, the number of emigrations of Croatian citizens who registered in settlements in different World countries in the same periods. Thus, the nodes of the network as defined in Section \ref{gendef} are in this case the Croatian settlements and the settlements of a particular country with which there is migration exchange, where the latter are contracted to single nodes (countries). The fact that we include on the one side (Croatian side) the precise settlements, while on the other side the countries as a whole, is because of of the lack of information on specific settlements for immigrants to Croatia prior to their move, or for Croatian citizens who registered in a different country after emigrating from Croatia within the same year. Nevertheless, this country-node contraction does not lead to a loss of generality when it comes to assessing the oveall network features, especially those traced in previous sections.

\begin{figure}[ht]
    \centering
    \includegraphics[width=\textwidth]{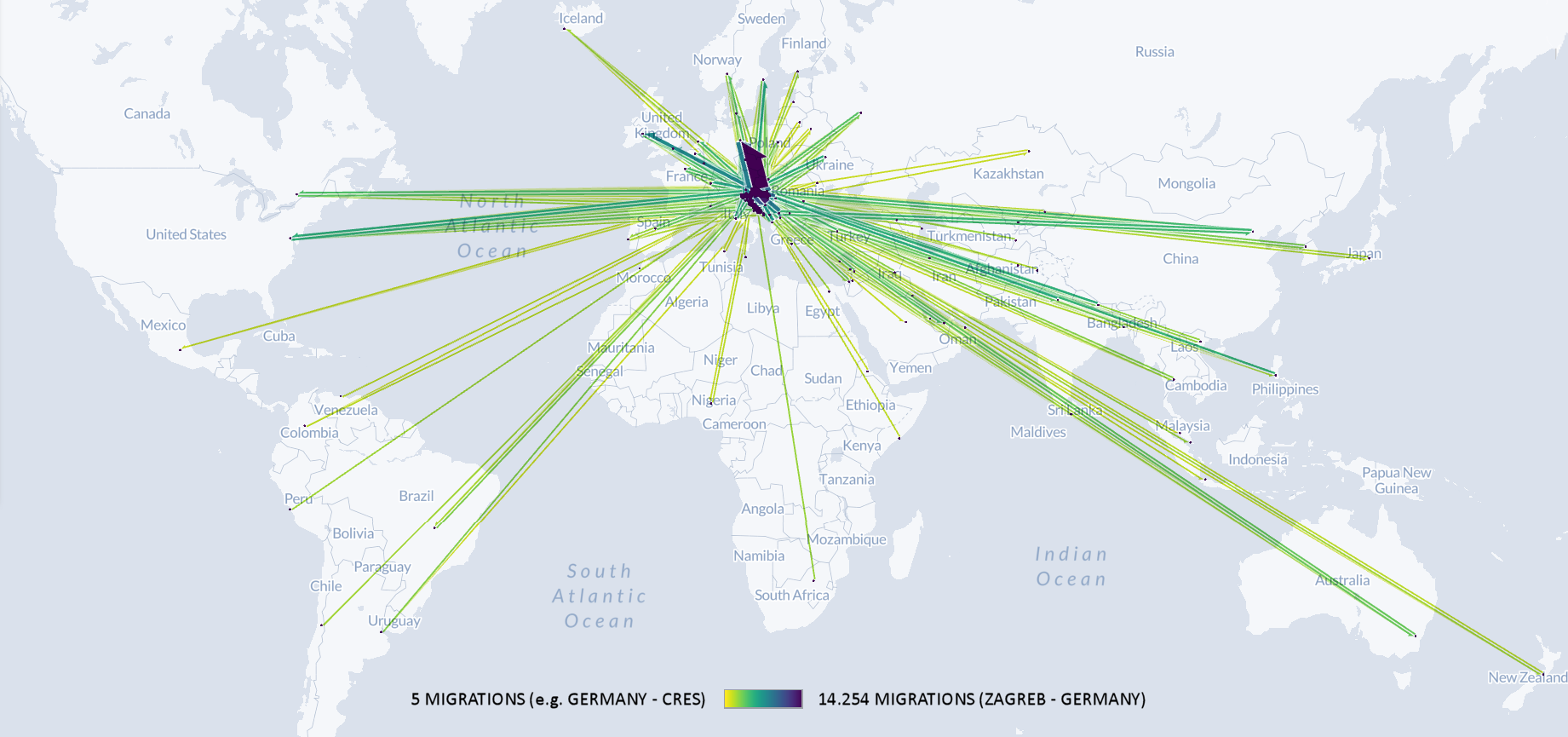}
    \caption{\textbf{International migration to/from Croatia, total 2016-2021}}
        \floatfoot{\small{Countries of immigration/emigration (citizenship countries of migrants) represented by country capitals. Link thickness reflects migration total sum of weights on links in the period. Interactive map available at \url{http://bit.ly/CROMIG}.}}
    \label{fig:5}
\end{figure}

In Figure \ref{fig:5} we present a screenshot of the analysed migration network which is available as a comprehensive interactive visualization at \url{https://bit.ly/CROMIG} (the $\mathcal{G^\prime}$ for this case analysis). The visualization may be particularly useful for Croatian policymakers, as they can quickly identify the links and nodes where migration exchange, particularly the higher levels of emigration, is more pronounced. We invite the readers, particularly the policymakers, to examine also the interactive visualization provided at \url{https://bit.ly/IMMCROCOUNTIES}, where emigration, immigration and total migration are shown at the level of the 23 counties in Croatia. Among other things, the map shows counties (regions) in Croatia that have alarmingly high emigration rates, summarizing the top emigration destinations. Geolocations (latitudes and longitudes of Croatian cities/municipalities and countries' capitals) are obtained from various free web services, and have further been validated by using Google Maps (\url{https://www.google.com/maps}).

In terms of the nodal metrics, considering that we examine a sub-network of the wider migration network (as we only observe migration exchange between Croatian settlements and settlements in other countries, excluding exchanges involving settlements from other world countries), it is feasible only to examine the weighted in- and weighted out- degree centrality of Croatian settlements per each year, which can be inferred from the interactive visualization (see separately immigration and emigration, or the total tab). We also cannot quantify migration on self loops as we do not have data on weights on self loops in settlements in other countries. What we generally can say for looping migration on the international level, is, to repeat the estimations referred to in the introduction, that internal migration is generally three times larger phenomenon than international migration \cite{McKenzie2022}. That means, essentially, that the sum of migration weights on self-loops at the intra-country level is three times higher than the sum of weights on all migration links connecting settlements from different countries, when global migration is observed. For Croatia, in 2018, the migration exchange with the world was 65.544 persons, while internal migration was 71.703 persons, but the latter number should be taken with reservations due to the proably understated intra-settlement migration figures (see the discussions in \cite{Pitoskietal2021b}).

\begin{figure}[ht]
    \centering
    \includegraphics[width=0.9\textwidth]{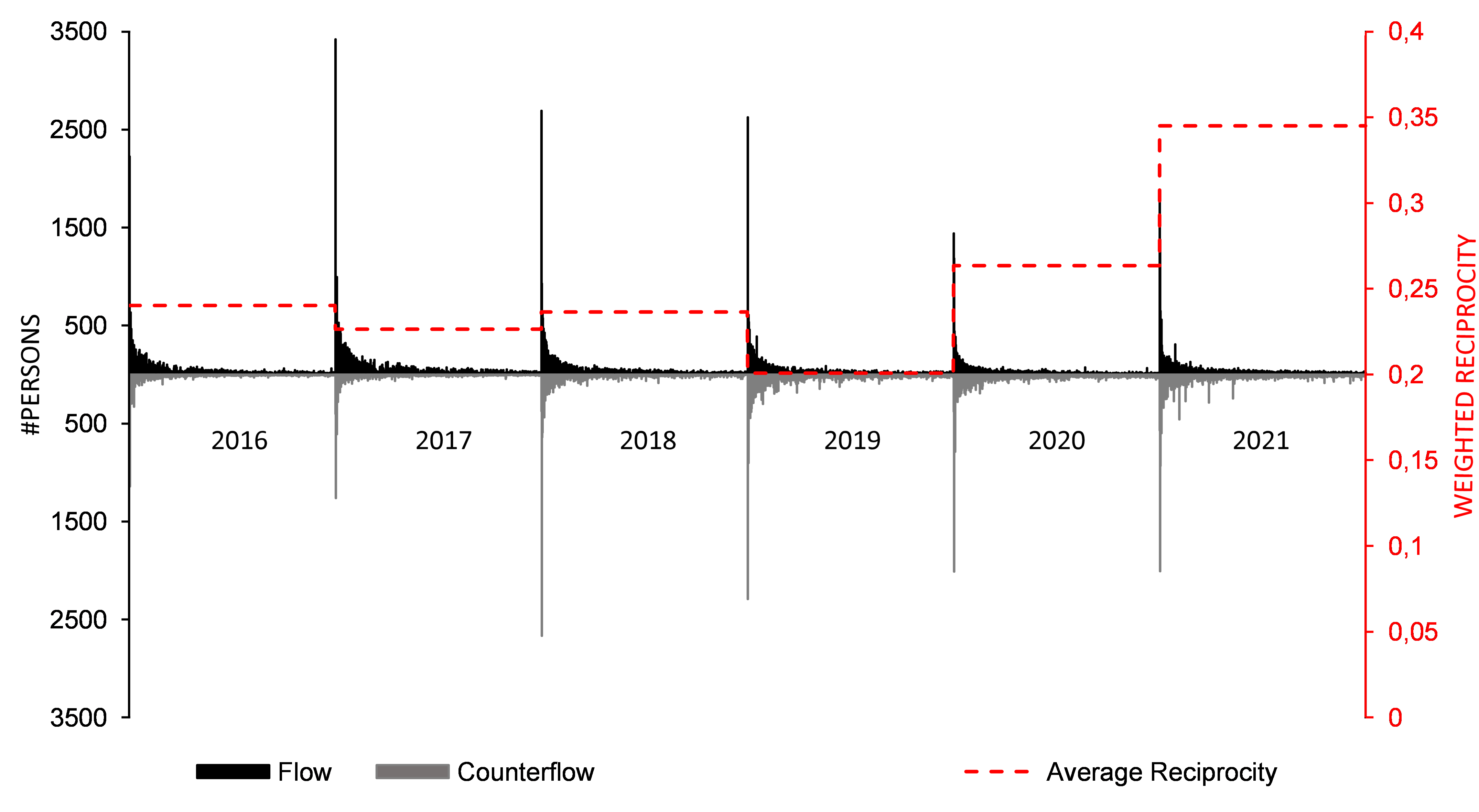}
    \caption{\textbf{Reciprocity of migration between Croatian settlements and World countries, 2016-2021}}
        \floatfoot{\small{In the chart above, each new period 
 is started with the link with highest total weight (flow) and counterweight (counterflow) in the whole period (Zagreb - Germany, Germany - Zagreb), proceeding in descending order in terms of total weight. See text for more explanation.}}
    \label{fig:6}
\end{figure}

What we can do reliably is assess weighted reciprocity, as another measure of particular interest, in line with its significance established from former observations at the city-scale and country-scale networks. In Figure \ref{fig:6} we provide a glance into the reciprocity on the first 830 links of 3582 links in total, on which there was significant international migration exchange (20\% of links whose weights cover for 80\% of all migration, in line with the ubiquitous Pareto's principle). The flows (in black) are weights on links going from Croatian settlements to other World countries, and the counterflows (in grey) are the weights on counterlinks returning from the same countries to the same Croatian settlements. For each year, the links are shown in the descending order in terms of the total weight realized on both link and counterlink in the whole period.

The reciprocity trend is such that from 2016 to 2018 it averaged around 23\%, then dropped slightly around 2019 to 20\% (probably related to the COVID crisis), and then rapidly increasing in the last few years, averaging at 35\%. Notably, reciprocity is even higher for the top 830 links (weights), where towards the end of the same period it climbed to about 49\%. We remind that this is the same value as the value calculated on the internal migration network the country as a whole (see Subsection \ref{Croint}). 

Conclusively, when the subnetwork of migration between human settlements in Croatia and other World countries is observed, and sticking to our manner of representation when referring to reciprocity, we can say that for every 100 people moving from city or municipality A in Croatia to some location in country B in the most recent year, on average we will find at least 35 people moving from the same country B to the same city/municipality A in the same period.

\section{Discussion}
\label{Sec4}

\begin{figure}[ht]
    \centering
    \includegraphics[width=\textwidth]{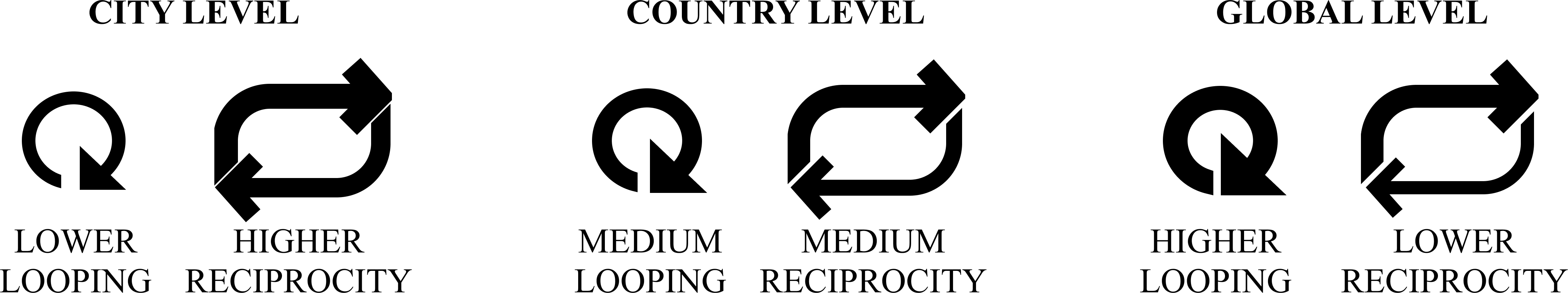}
    \caption{\textbf{Characteristic migration network patterns by geographical scale}}
    \label{fig:7}
\end{figure}

As the begining of our discussion we can summarize the conclusions of our analyses when it comes to general features of migration networks, aided by Figure \ref{fig:7}, as follows. 

Based on the cases investigated (to the extent that we can claim geographical, historical, and external validity in general) we show that migration networks are characterized by the main traits of pronounced self loops and pronounced reciprocity on all geographical scales. These traits vary across different geographical scales in the following way; at the narrower geographical scales (intra-city migration networks) we may typically find lower migration from one to the same location against a higher share of reciprocal migration between any two different locations (locations here being represented by city districts). At the intermediate geographical scales (intra-country migration networks) one will find both of these features to be about of the same strength; about half of all migration will be from one to the same location, while the other half will be the reciprocal exchange between the ``established'' location dyads (locations here being represented by country municipalities or cities). At the global scale, one may expect to see higher within-location migration against between-location migration (locations again being represented by municipalities/cities), although reciprocal flows between established location dyads are still very much pronounced. 

Now what does this mean for the analysis of migration factors which underlie these network formations? As the literature review of the factors of realized migration summarized in Section \ref{Sec2} shows, in literature we will typically find the following generalized regression model through which we infer on the migration factors' influence:

\begin{equation}
M_{i,j} = \beta_0 + \sum_{k=1}^{n} \beta_k F_{i,k} + \sum_{l=1}^{n} \beta_l F_{j,l} + \sum_{m=1}^{n} \beta_m F_{i,j,m} + \varepsilon
\end{equation}
\label{eq:1}

Where:
\begin{itemize}
    \item $M_{i,j}$ represents the dependent variable denoting migration from origin $i$ to destination $j$,
    \item $F_{i,k}$ and $F_{j,l}$ represent origin-specific and destination-specific migration factors, respectively,
    \item $F_{i,j,m}$ represent an origin-destination link (intervening) migration factor, and 
    \item $\beta_0$, $\beta_k$, $\beta_l$, $\beta_m$, and $\varepsilon$ represent, respectively, the intercept, origin-specific factor coefficients, destination-specific factor coefficients, link factor coefficients and the error term.
\end{itemize}

Note that this generalized form varies across migration factor studies in numerous ways; for example, the regression-analysis based paper on reasons for internal migration in Austria that we touched upon earlier in the paper \cite{Jestletal2022} takes only the emigration rate as the dependent variable in the model. For the numerous variable operationalization variants, refer to the work cited in Section \ref{Sec2}. Still, this is a good general representation on the most represented general model as the proof technique for determining migration factor effects in literature.

Now, what we conclude to be problematic with this kind of modelling, based on the conclusions that we deliver from the network science approach, is that this typical model representation (and its variations) does not take into account any of the pronounced network characteristics discovered in the former sections; neither high weights on self loops, nor high reciprocity of origin-destination links. It is arguably the case that, if simultaneously within the set of values for the dependent variable (migration) and in the set of values for the independent or other-type variables (migration factors) we include both those values that pertain to origin-destination flows/factors and those that pertain to destination-origin flows/factors, the models become unreliable. In addition, if we know that looping migration is such a sizable phenomenon that occurs simultaneously with the sizable reciprocity, why is it that almost none of the models includes this aspect into the observation?

We can propose several suggestions on how to enhance these general models, but also network models, and integrate two fields in general; from those relatively obvious (and perhaps already deployed in some rare cases), to those novel and relatively demanding. The first and most obvious way on how to update the models is to take the non-reciprocated weights of migration flow on links as values of the dependent variable, which would deliver factors that underlie migration between locations. Morover, the factors that determine migration within locations (looping migration), must be introduced in the same models (as $M_{i,i}$ and $F_{i,i}$) added subsequently to each side of equation, as a form of a robustness check. Inclusion of self-loop migration applies both to regression analyses and network analyses, as network scientists' typical way of abstracting migration networks, as to statisticians up to date, has been simply a reduction to between-location migration (i.e. ignoring the migration on self-loops).

The second way on how to improve regression models is to incorporate network indicators as predictors or control variables in the models. These can be centrality indicators, or other-than-local indicators such as clustering or assortativity coefficient. Rare studies have been done that use that approach in a complex and innovative way, also upgrading from standard regression modelling (see, e.g. \cite{WINDZIO201820, Windzioetal2019}), yet these are again limited to between-location migration. The models must simultaneously include centralities (e.g. node strength) or other coefficients that include within-location migration. However, as it has been clearly shown in \cite{Pitoskietal2021a, Pitoskietal2021b,}, most network measures fail with high levels of reciprocity (and looping) in the abstracted static (weighted) networks and the approach in general becomes almost meaningless due to the reduced feasibility of network measure applications. This feasibility is particularly lower for eigencentrality metrics and other indirect-connectivity based measures such as PageRank, HITS centrality, transitivity or assortativity coefficients, while community detection is also possibly biased.

All these obstacles, on both the statistics and network science side, can be corrected with space-time or dynamic network abstractions before running any models; the issue we already touched upon in Section \ref{Sec2}. This means observing nodes as space-time positions (of a person or of a ``batch'' of people) moving from one to the same or different location in time. For the generalized model of calculation of centrality in such abstracted network, which centrality can actually be used as a dependent or other kind of variable in a regression model, see \cite{Pitoskietal2023a}. Suitable data for space-time migration network abstractions is needed to make this possible, and more measures need to be upgraded to be applied on such abstractions. The recently launched database for the Netherlands (``the Dutch population Statistics Netherlands Microdata Catalogue'') which also contains space-time positions of people over years along with other personal attributes (see, for example, \cite{Bokanyietal2023}), enables creating such space-time network abstractions before measure applications. However, dynamic networks measures development or adjustation, at the same time, need to get very intense, as this is a part of network science research which is severely under-developed \cite{Pitoskietal2023a}.

Ultimately, drivers of human migration may be incorporated directly in network science indicators, where network links are weighted by the values for migration factors in multilayered, and preferably dynamic (space-time) networks, on which subsequently the network measures are applied. For example, income differential between any two locations in the migration network can determine the link weight in one layer. Another layer could be the kilometer distance between locations. Language proximity or contiguity could determine the third or forth, binary network layers, and so on. Node centralities and other indicators calculated on such networks (calculated for each layer separately as well as cumulatively) could be used as dependent/independent/other-type variables in regression analyses.

All the above suggestions, however, require a proof of concept based on a real case analysis, which we are to undertake in our future work. Nevertheless, we take this study to be of good value in terms of revealing the gaps within and between Migration Studies and Network Science applied to human migration. Moreover, it provides a comprehensive overview of patterns of human migration, which is undoubtedly very useful for the policymakers of analysed countries, Austria and Croatia, but also wider. At this point, while migration is perhaps still a manageable phenomenon, the policymakers may incorporate the knowledge on migration patterns to finally start building policies that will handle migration which is destined to intensify, and have an impact on all societal aspects, in the very near future.

\newpage

\section*{Declarations}

\subsection*{Ethical approval}
Not applicable.

\subsection*{Competing interests}
Not applicable.

\subsection*{Authors' contributions}
Conceptualization: D.P.; methodology: D.P.; software: D.P; validation: D.P., H.S. and A.M.; formal analysis: D.P.; investigation: D.P.; resources: D.P., H.S.; data curation: D.P.; writing–original draft preparation: D.P.; writing–review and editing: D.P., H.S. and A.M.; visualization: D.P.; supervision: H.S. and A.M.; project administration: D.P., H.S., and A.M.; funding acquisition: D.P. All authors have read and agreed to the published version of the manuscript.

\subsection*{Funding} This research was supported by the Young Universities for the Future of Europe Alliance (YUFE, \url{https://yufe.eu/}), as part of the Postdoctoral Programme on the ``Citizens' Wellbeing'' (call year 2021).

\subsection*{Availability of data and materials} All datasets generated and/or analysed in this study are available via the web links and references provided in the manuscript. All URLs and DOIs provided have been (re)accessed on 28th August 2023.

\subsection*{Acknowledgements} Authors would like to express their special thanks to the fellow members of the Laboratory for Semantic Technologies at the Faculty of Informatics and Digital Technologies at University of Rijeka (FIDIT), the Laboratory for Complex Networks at the Centre for Artificial Intelligence and Cybersecurity, University of Rijeka (AIRI), the Dept. of Living Conditions and Social Cohesion of the Central Bureau of Statistics (CBS), office Heerlen, the Netherlands, and the Dept. of Political Science of Faculty of Arts and Social Sciences at Maastricht University (FASoS), for their comments that helped to improve the manuscript.

\printbibliography

@book{clement2021groundswell,
  title = {Groundswell Part 2: Acting on Internal Climate Migration},
  author = {Clement, Viviane and Rigaud, Kanta Kumari and de Sherbinin, Alex and Jones, Bryan and Adamo, Susana and Schewe, Jacob and Sadiq, Nian and Shabahat, Elham},
  year = {2021},
  publisher = {World Bank},
  url = {http://hdl.handle.net/10986/36248},
  license = {{CC BY 3.0 IGO}}
}

@book{UNPF2023,
   author = "United Nations Population Fund",   
    title = "State of World Population 2023",
   url = "https://www.un-ilibrary.org/content/books/9789210027137"
}

@article{Pitoskietal2023a,
author = {Pitoski, Dino and Babić, Karlo and Meštrović, Ana},
  title = {A new measure of node centrality on schedule-based space-time networks for the designation of spread potential},
  year = {2023},
  journal = {submitted for publication to Scientific Reports, preprint available at Research Square},
  url ={https://assets.researchsquare.com/files/rs-2474713/v1/e026055f986e1ff332210102.pdf?c=1674017533}
}

@inproceedings{Pitoskietal2021c,
	address = {Wiesbaden},
	title = {Human migration as a complex network: appropriate abstraction, and the feasibility of {Network} {Science} tools},
	booktitle = {Data Science–Analytics and Applications: Proceedings of the 3rd International Data Science Conference iDSC2020},
	publisher = {Springer Fachmedien Wiesbaden},
	author = {Pitoski, Dino and Lampoltshammer, Thomas J. and Parycek, Peter},
	editor = {Haber, Peter and Lampoltshammer, Thomas and Mayr, Manfred and Plankensteiner, Kathrin},
	year = {2021},
doi={10.1007/978-3-658-32182-6_17}
}

@article{Pitoskietal2021a,
author = {Pitoski, Dino and Lampoltshammer, Thomas J. and Parycek, Peter},
title = {Network Analysis of Internal Migration in Austria},
year = {2021},
journal = {Digital Governance: Research and Practice},
volume = {2},
number = {3},
doi = {10.1145/3447539},
journal = {Digit. Gov.: Res. Pract.}
}

@article{Pitoskietal2021b,
  author={Pitoski, Dino and Lampoltshammer, Thomas Josef and Parycek, Peter},
 title={Network analysis of internal migration in Croatia},
  journal={Computational Social Networks},
  volume={8},
  number={10},
  year={2021},
  doi={10.1186/s40649-021-00093-0}
}

@misc{authorsown2,
	author = {Pitoski, Dino, and Lampoltshammer, Thomas J., and Parycek, Peter},
	title  = {Network Analysis of Internal Migration in Croatia - Supplementary material. figshare. Dataset},
	year={2020},
	url = {https://doi.org/10.6084/m9.figshare.12497177}
}

@inproceedings{Pitoskietal2023b,
  author = {Pitoski, Dino and Beliga, Slobodan and Meštrović, Ana},
  title = {First Insight into Social Media User Sentiment Spreading Potential to Enhance the Conceptual Model for Disinformation Detection},
  booktitle = {Data Science–Analytics and Applications: Proceedings of the 5th International Data Science Conference - iDSC2023},
  year = {2023},
  note = {Forthcoming},
}

@misc{WBmigdata,
	author = {{The World Bank}},
	title  = {Migration and Remittances Data},
	url = {https://www.worldbank.org/en/topic/migrationremittancesdiasporaissues/brief/migration-remittances-data}
}

@ARTICLE{Squartini2013-xw,
   title    = "Reciprocity of weighted networks",
   author   = "Squartini, Tiziano and Picciolo, Francesco and Ruzzenenti, Franco and Garlaschelli, Diego",
   journal  = "Sci. Rep.",
   volume   =  {3},   
number   =  {2729},
   year     =  2013,
  doi={10.1038/srep02729}
 }

@ARTICLE{Blondel2008-ug,
   title     = "Fast unfolding of communities in large networks",
   author    = "Blondel, Vincent D. and Guillaume, Jean-Loup and Lambiotte,
                Renaud and Lefebvre, Etienne",
   journal   = "J. Stat. Mech.",
   publisher = "IOP Publishing",
volume = {2008},
number = {10},
   year      =  2008,
   language  = "en",
doi={10.1088/1742-5468/2008/10/P10008}
 }

@article{Schon2021,
      author = "Justin Schon",
      title = "Migration Causes, Patterns, and Consequences: Contributions of Location Networks",
      year = "2021",
journal = {Oxford Research Encyclopedias: Politics},
      doi = {10.1093/acrefore/9780190228637.013.1962}
}

@misc{CBS2018,
  title={Migratiemotieven van immigranten met een buitenlandse nationaliteit (VRLMIGMOTBUS)},
  author={{Central Bureau of Statistics of the Netherlands}},
  year={2018},
  url={https://www.cbs.nl/onze-diensten/maatwerk-en-microdata/microdata-zelf-onderzoek-doen/microdatabestanden/vrlmigmotbus-migratiemotieven},
}

@article{Schmeets2019,
  title={Migranten vertrekken eerder uit Nederland},
  author={Schmeets, Hans},
  journal={Economisch Statistische Berichten},
  volume={104},
  number={4475},
  year={2019},
url={https://esb.nu/migranten-vertrekken-eerder-uit-nederland/}
}

@inproceedings{Aleskerovetal2020,
  title={New Centrality Measures in Networks and their Applications to the International Trade and Migration Networks},
  author={Aleskerov, Fuad and Roman, Andrey and Rezyapova, Anzelika and Yakuba, Vyacheslav},
  booktitle={2020 28th International Symposium on Modeling, Analysis, and Simulation of Computer and Telecommunication Systems (MASCOTS)},
  year={2020},
  organization={IEEE},
  doi={10.1109/mascots50786.2020.9285957},
}

@article{Wangetal2020,
  title={Complex Network of Scientific Talent Migration in Discrete Dynamics from 2001 to 2013},
  author={Wang, Yinqiu and Luo, Hui and Shi, Yunyan},
  journal={Discrete Dynamics in Nature and Society},
  volume={2020},
  number={9248983},
  year={2020},
  publisher={Hindawi},
  doi={10.1155/2020/9248983},
}

@article{Bonaccorsietal2019,
  title={Country Centrality in the International Multiplex Network},
  author={Bonaccorsi, Giovanni and Riccaboni, Massimo and Fagiolo, Giorgio and Santoni, Gianluca},
  journal={Applied Network Science},
  year={2019},
  volume={4},
  number={126},
  doi={10.1007/s41109-019-0207-3},
}

@article{Windzioetal2019,
  title={A Network Analysis of Intra-EU Migration Flows: How Regulatory Policies, Economic Inequalities and the Network-Topology Shape the Intra-EU Migration Space},
  author={Windzio, Michael and Teney, Céline and Lenkewitz, Sven},
  journal={Journal of Ethnic and Migration Studies},
  year={2019},
 volume = {47},
number = {5},
  doi={10.1080/1369183X.2019.1643229},
}

@article{WINDZIO201820,
title = {The network of global migration 1990–2013: Using ERGMs to test theories of migration between countries},
journal = {Social Networks},
volume = {53},
year = {2018},
note = {The missing link: Social network analysis in migration and transnationalism},
doi = {https://doi.org/10.1016/j.socnet.2017.08.006},
author = {Michael Windzio}
}

@article{Bokanyietal2023,
  title={The anatomy of a population-scale social network},
  author={Bok{\'a}nyi, Eszter and Heemskerk, Eelke M. and Takes, Frank W.},
  journal={Scientific Reports},
  year={2023},
  volume={13},
  number={1},
  doi={10.1038/s41598-023-36324-9}
}

@article{Rochaetal2022,
  title={The Global Migration Network of Sex-Workers},
  author={Rocha, Luis E. C. and Holme, Petter and Linhares, Claudio D. G.},
  journal={Journal of Computational Social Science},
  volume={5},
  year={2022},
  publisher={Springer},
  doi={10.1007/s42001-021-00156-2},
}

@article{GursoyandBadur2022,
  title={Investigating Internal Migration with Network Analysis and Latent Space Representations: An Application to Turkey},
  author={Gürsoy, Furkan and Badur, Bertan},
  journal={Social Network Analysis and Mining},
  volume={12},
    number={150},
  year={2022},
  publisher={Springer},
  doi={10.1007/s13278-022-00974-w},
}

@article{PoratandBenguigui2021,
title = {Global migration topology analysis and modeling of directed flow network 2006–2010},
journal = {Physica A: Statistical Mechanics and its Applications},
volume = {561},
number = {125210},
year = {2021},
doi = {https://doi.org/10.1016/j.physa.2020.125210},
author = {Idan Porat and Lucien Benguigui}
}

@article{Akbari2021,
  title={Exploratory Social-Spatial Network Analysis of Global Migration Structure},
  author={Akbari, Hossein},
  journal={Social Networks},
  volume={64},
  year={2021},
  doi={10.1016/j.socnet.2020.09.007},
}

@article{Mourao2020,
  title={Footsteps in the Sand: Studying Refugee Paths since 2005 through a Network Analysis of 205 Territories},
  author={Mourao, Paulo Reis},
  journal={Quality \& Quantity},
volume = {55},
  year={2020},
  doi={10.1007/s11135-020-01014-5},
}

@article{GursoyandBadur2021,
    author = {Gürsoy, Furkan and Badur, Bertan},
    title = "{Extracting the signed backbone of intrinsically dense weighted networks}",
    journal = {Journal of Complex Networks},
    volume = {9},
    number = {5},
    year = {2021},
    doi = {10.1093/comnet/cnab019},
}

@article{Tjaden2021,
  title={Measuring Migration 2.0: A Review of Digital Data Sources},
  author={Tjaden, Jasper},
  journal={Comparative Migration Studies},
  year={2021},
  volume={9},
  number={59},
  doi={10.1186/s40878-021-00273-x},
}

@article{CarvahloandChalresEdwards2020,
author = {Carvalho, Rodrigo Coelho de and Charles-Edwards, Elin},
title = {The evolution of spatial networks of migration in Brazil between 1980 and 2010},
journal = {Population, Space and Place},
volume = {26},
number = {e2332},
doi = {https://doi.org/10.1002/psp.2332},
year = {2020}
}

@article{Chenetal2021,
author = {Yu-wang Chen and Lei Ni and Luis Ospina-Forero},
title ={Visualising internal migration flows across local authorities in England and Wales},
journal = {Environment and Planning A: Economy and Space},
volume = {53},
number = {4},
year = {2021},
doi = {10.1177/0308518X20968568},
}

@article{Lee1966,
 doi = {10.2307/2060063},
 author = {Everett S. Lee},
 journal = {Demography},
 number = {1},
 publisher = {Springer},
 title = {A Theory of Migration},
 volume = {3},
 year = {1966}
}

@article{Mabogunje1970,
author = {Mabogunje, Akin L.},
title = {Systems Approach to a Theory of Rural-Urban Migration},
journal = {Geographical Analysis},
volume = {2},
doi = {10.1111/j.1538-4632.1970.tb00140.x},
year = {1970}
}

@article{Zelinsky1971,
 doi = {10.2307/213996},
 author = {Wilbur Zelinsky},
 journal = {Geographical Review},
 number = {2},
 publisher = {[American Geographical Society, Wiley]},
 title = {The Hypothesis of the Mobility Transition},
 volume = {61},
 year = {1971}
}

@book{Skeldon1990,
  author = {Ronald Skeldon},
  title = {Population Mobility in Developing Countries: A Reinterpretation},
  year = {1990},
  publisher = {Belhaven Press},
  address = {London}
}

@article{HarrisandTodaro1970,
  author = {John R. Harris AND Michael P. Todaro},
  title = {Migration, Unemployment and Development: A Two-Sector Analysis},
  journal = {American Economic Review},
  volume = {60},
  number = {1},
  year = {1970},
url={http://www.jstor.org/stable/1807860}
}

@book{Piore1979,
  author = {Michael J. Piore},
  title = {Birds of Passage: Migrant Labor and Industrial Societies},
  year = {1979},
  publisher = {Cambridge University Press},
  address = {Cambridge}
}

@book{Stark1978,
  author = {Oded Stark},
  title = {Economic-Demographic Interactions in Agricultural Development: The Case of Rural-to-Urban Migration},
  year = {1978},
  publisher = {FAO},
  address = {Rome}
}

@book{Stark1991,
  author = {Oded Stark},
  title = {The Migration of Labor},
  year = {1991},
  publisher = {Blackwell},
  address = {Cambridge and Oxford}
}

@article{Massey1990,
  author = {Douglas S. Massey},
  title = {Social Structure, Household Strategies, and the Cumulative Causation of Migration},
  journal = {Population Index},
  volume = {56},
  number = {1},
  year = {1990},
doi={10.2307/3644186}
}

@article{deHaas2021,
  author = {Hein de Haas},
  title = {A Theory of Migration: The Aspirations-Capabilities Framework},
  journal = {Comparative Migration Studies},
  volume = {9},
  number = {8},
  year = {2021},
  doi = {10.1186/s40878-020-00210-4}
}

@misc{Aslanyetal2021,
  author = {Aslany, Maryam and Jørgen Carling and Mathilde Bålsrud Mjelva and Tone Sommerfelt},
  title = {Systematic review of determinants of migration aspirations. QuantMig deliverable, 2.2},
  howpublished = {The Peace Research Institute Oslo (PRIO)},
  url = {https://www.prio.org/publications/12613},
  year = {2021},
}

@misc{McKenzie2022,
  author = {McKenzie, David},
  title = {Fears and Tears: Should More People Be Moving within and from Developing Countries, and What Stops This Movement?},
  howpublished = {The World Bank. Policy Research Working Paper No. 10128},
  year = {2022},
  url = {http://hdl.handle.net/10986/37759},
}

@misc{dzs2023,
  author = "{Central Bureau of Statistics Croatia}",
  title = "Objavljeni konačni rezultati Popisa 2021",
url={https://dzs.gov.hr/en/vijesti/objavljeni-konacni-rezultati-popisa-2021/1270},
}

@article{Pisarevskayaetal2020,
author = {Pisarevskaya, Asya and Levy, Nathan and Scholten, Peter and Jansen, Joost},
title = {Mapping migration studies: An empirical analysis of the coming of age of a research field},
journal = {Migration Studies},
volume = {8},
year = {2020},
doi = {10.1093/migration/mnz031}
}

@Article{Pitoskietal2021d,
AUTHOR = {Pitoski, Dino and Lampoltshammer, Thomas J. and Parycek, Peter},
TITLE = {Drivers of Human Migration: A Review of Scientific Evidence},
JOURNAL = {Social Sciences},
VOLUME = {10},
YEAR = {2021},
NUMBER = {1},
ARTICLE-NUMBER = {21},
DOI = {10.3390/socsci10010021}
}

@misc{Magistrat-Stadt-Wien-2020,
  author       = {{Magistrat der Stadt Wienn - Stadt Wien Wirtschaft, Arbeit und Statistik}},
  title        = {Statistisches Jahrbuch der Stadt Wien 2019},
  year         = {2020},
  howpublished = {\url{https://web.archive.org/web/20200701040527/https://www.wien.gv.at/statistik/pdf/jahrbuch-2019.pdf}},
}

@misc{migrationgvat_geography_population,
  title = "Geography and Population",
  author = "Federal Ministry of Labour and Economics of Austria and Federal Ministry of the Interior of Austria",
  year = "2003",
  howpublished = "\url{https://www.migration.gv.at/en/living-and-working-in-austria/austria-at-a-glance/geography-and-population/}",
}

@article{Jestletal2022,
  title={Cannot keep up with the Joneses: how relative deprivation pushes internal migration in Austria},
  author={Jestl, Stefan and Moser, Mathias and Raggl, Anna-Katharina},
  journal={International Journal of Social Economics},
  volume={49},
  number={2},
  year={2022},
  doi={10.1108/IJSE-03-2021-0181}
}

\end{document}